\documentclass[12pt]{JHEP3}
\usepackage{graphicx}
\usepackage{dcolumn}  
\usepackage{bm}    
\usepackage{amssymb} 
\usepackage{amsmath,bm}
\usepackage{amsfonts}    
\usepackage{slashed}  
\usepackage[mathscr]{euscript}
\title{Black holes  in  higher spin supergravity}
\author{
Shouvik Datta,  Justin R. David   \\
 Centre for High Energy Physics,
Indian Institute of Science,\\ C.V. Raman Avenue, Bangalore 560012, India. \\
\email{shouvik, justin@cts.iisc.ernet.in}\\
}
\abstract{ We study 
black hole solutions in
 Chern-Simons higher spin supergravity based on the superalgebra $sl(3|2)$. 
These black hole solutions have  a $U(1)$ gauge field and a spin $2$ hair in addition to 
the spin $3$ hair.  These additional fields correspond to the  R-symmetry charges of the supergroup
 $sl(3|2)$. Using the relation between the bulk field equations and the Ward identities of 
a CFT with ${\cal N}=2$  super-${\cal W}_3$ 
symmetry,  we identify the bulk charges and chemical potentials
with those of the boundary CFT.  
From these identifications we see that a suitable set of  variables to study this black hole is in terms of 
the charges present in  three decoupled bosonic sub-algebras of 
the ${\cal N}=2$ super-${\cal W}_3$ algebra. 
The entropy and the partition function  of these R-charged black holes are then evaluated in terms of the  charges of the bulk theory as well as in terms of its chemical potentials. 
We  then compute the partition function in the dual CFT and find exact agreement 
with the bulk partition function. 
}

\begin{document}
\def\G{\Gamma}
\def\lb{\left(}
\def\rb{\right)}
\def\k{\kappa}
\def\nn{\nonumber}
\def\pU{U^+}
\def\mU{U^-}
\def\pmU{U^\pm}
\def\pG{G^+}
\def\mG{G^-}
\def\pmG{G^\pm}
\def\nn{\nonumber}
\def\cJ{\mathcal{J}}
\def\cV{\mathcal{V}}
\def\cL{\mathcal{L}}
\def\cW{\mathcal{W}}
\def\pd{\partial}
\def\bz{{\bar z}}

\section{Introduction}
The study of higher spin theories $(s\geq2)$ in Anti-de Sitter spaces have been the focus of many 
recent works (see \cite{Vasiliev:2003ev,Gaberdiel:2012uj} for comprehensive
reviews). These theories offer toy models of  AdS/CFT 
with enhanced symmetry and without the complications of the infinite tower of massive 
string excitations.  Higher spin theories in $3$ dimensional Anti-de Sitter space are 
easier to formulate in terms of Chern-Simons theory which renders them more tractable
\cite{Blencowe:1988gj,Henneaux:2010xg,Henneaux:2012ny,Campoleoni:2010zq}. 
In three dimensions it is also possible to 
consistently truncate  to a finite set of higher spin fields. 
There are explicit proposals for the CFT  duals of Vasiliev like theories in terms of minimal
models for cases with and without supersymmetry \cite{Gaberdiel:2010pz, Creutzig:2011fe}. 

Classical solutions in higher spin theories in $3$ dimensional Anti-de Sitter space 
are easy to construct (see \cite{Ammon:2012wc} for a recent review). 
This is because in terms of the Chern-Simons formulation they 
correspond to a flat connection. Black holes and conical defect solutions
constructed  in these theories have been used to study the spectrum, nature
of singularities with enhanced higher spin symmetry and the holographic renormalization
group \cite{Castro:2011iw,Castro:2011fm,Ammon:2011nk}. 
 Recently the conditions under which a  classical solution 
is supersymmetric in a Chern-Simons higher spin theory was provided in \cite{Tan:2012xi,Datta:2012km}. 
Supersymmetric classical solutions should correspond to chiral primaries in the
boundary CFT and a recent  check of this fact  has been done in \cite{Hikida:2012eu}.

Conformal field theory  in 2 dimensions with enhanced supersymmetry
usually admits redefinitions of the currents so that the bosonic 
sub-algebras mutually commute.  
Consider the simple case of a CFT with ${\cal N}=2$ supersymmetry. 
The bosonic part of this algebra consists of the stress tensor $T$ and the $U(1)$ current
$J$.  Redefining the stress tensor as 
\begin{equation}
 \hat T=  T- \frac{3}{2 c} J^2 
\end{equation}
ensures that the new Virasoro algebra commutes with the $U(1)$ current. 
This redefinition has an important consequence for a BTZ black hole carrying $U(1)$ charge. 
A bulk theory with ${\cal N}=2$ asymptotic algebra is a Chern-Simons theory 
based on the supergroup $sl(2|1)$. 
The bosonic spectrum of this theory consists of the graviton and a Chern-Simons $U(1)$ gauge field. 
Consider the BTZ black hole along with the $U(1)$ gauge field. 
The metric of the black hole is unaffected by the $U(1)$ gauge field. Thus the entropy 
is independent of  the presence of $U(1)$ field.
However in the AdS/CFT correspondence,  it is $-\hat T$ which is identified with the 
stress tensor of the bulk
 \cite{Schwimmer:1986mf,Henneaux:1999ib,Maldacena:2000dr}.
This shift 
is due to the extra energy carried by the gauge field.  The  Cardy-like formula for the entropy now
is given by\footnote{See \cite{Kraus:2006wn} for a derivation.}
\begin{equation}
 S =  2\pi \sqrt{ \frac{c}{6} \left(  L_0  - \frac{3}{2c} J_0^2 \right) }
\end{equation}
where $L_0$ and $J_0$ are the zero modes of the stress tensor and the $U(1)$ current
of the CFT. 
We wish to study this phenomena in the presence of higher spin fields. A  natural and consistent
framework to introduce an additional $U(1)$ in Vasiliev theories is to study 
supersymmetric Chern-Simons theories based on the supergroup $sl(N|N-1)$. 
These theories always contain a $U(1)$ as a R-symmetry. 
The second goal in this paper is to  reproduce the 
partition function of black holes in  higher spin supersymmetric theories from the CFT. 

Our set up is the Chern-Simons theory based on the supergroup $sl(3|2)$ whose 
bosonic field content include  the graviton and  a spin 3 field. 
There is  an additional spin 2 field 
and a $U(1)$ gauge field corresponding to the R-symmetry part of the supergroup $sl(3|2)$. 
To discover how the charges in the boundary are affected due to the presence of the $U(1)$ field we 
use the correspondence of the bulk equations of motion to the Ward identities 
of the CFT \cite{Gutperle:2011kf}. 
In this case the Ward identities arise from the semi-classical operator product 
expansions  of the
the ${\cal N}=2$ super-${\cal W}_3$ which is the
asymptotic algebra of the $sl(3|2)$ theory\footnote{Asymptotic
algebras of higher spin super Chern-Simons theories were studied in 
\cite{Henneaux:2012ny,Tan:2012xi}.}. 
We see that indeed  the stress tensor is 
shifted precisely as expected. However, there is also a shift of the 
spin $3$ charge which is proportional to the product of the spin $2$ R-charge and the 
$U(1)$ charge.  The precise shifts are provided in Table 1.
 The bosonic part of the ${\cal N}=2$ super-${\cal W}_3$ algebra can 
be written into $3$ mutually commuting algebras given by 
\begin{equation} \label{decoup}
\mathcal{N}=2\text{ super-}{\cal W}_3 \supset {\cal W}^+ \oplus {\cal W}^- \oplus u(1)
\end{equation} 
where ${\cal W}^+$ is a Virasoro algebra and  ${\cal W}^-$ is  the bosonic ${\cal W}_3$ 
algebra \cite{Romans:1991wi}. 
From these shifts found in table 1,   we see that a natural linear combination of the 
 bulk charges are directly identified with the 
charges in the mutually commuting bosonic algebras. 
We then construct a black hole solution carrying these charges and evaluate its
partition function and  entropy. This is done
 using the integrability conditions satisfied by the holonomy equations.
The  partition function and the entropy  are written both in terms of the charges and the 
corresponding chemical potentials.  
We see that the black hole entropy can be written in terms of contributions from the decoupled
CFTs.  The contribution from the ${\cal W}^+$ part is due to a Virasoro algebra and the entropy
from the CFT can be understood in terms of the Cardy formula. 
The contribution  from the ${\cal W}^-$ part to the partition function with higher spin 
charge has been recently computed in \cite{Gaberdiel:2012yb}. 
Appealing to this result we show that the 
entropy of the black hole in the higher spin super-Chern-Simons  
theory can be exactly reproduced form the  CFT. 
We then discuss the implications of these observations for black holes in 
theories dual to the Kazama-Suzuki models of \cite{Creutzig:2011fe}. 

The organization of the paper is as follows: Section 2 reviews the generalities 
of higher spin super Chern-Simons theory and then details the $sl(3|2)$ algebra. 
We then write down the general form of the Chern-Simons connection which we will 
be interested in. 
In Section 3 the correspondence of the bulk equations of motion with the 
semi-classical Ward identities of the ${\cal N}=2$ super-${\cal W}_3$ algebra is used 
to obtain the relationship between the bulk charges and chemical potentials with 
that of the boundary theory.  From these we see that a natural linear combination
of  bulk charges are 
directly identified with the charges of the decoupled CFTs given in (\ref{decoup}). 
Section 4 discusses the construction of the higher spin black hole with 
spin 3 field carrying  the spin 2 and $U(1)$  R-charges.  
It then demonstrates the integrability conditions are satisfied by the 
holonomy conditions. This implies  the existence 
of consistent thermodynamic description of the black hole. 
We then evaluate the entropy  and the partition function of the black hole solution both 
in  terms of charges and the chemical potentials. 
We then show that the partition function can be reproduced exactly from the CFT. 
 The conditions for which the black hole is supersymmetric is also analyzed in this section.
We then discuss the implications of these observations for  black holes in
supersymmetric higher spin theories based on the supergroup $shs[\lambda]$.  
Section 6 contains the conclusions. 
Appendix A contains the details of the semiclassical OPEs of the 
${\cal N}=2$ super ${\cal W}_3$ algebra. Appendix B demonstrates the equivalence of the 
entropy written in terms of charges to that written in terms of the chemical potentials.

\textbf{Note added:} While this manuscript was in preparation we received \cite{Chen:2013oxa} which has some overlap with this paper.

\section{Constructing higher spin black holes}

\subsection{Higher spin Chern-Simons supergravity}
 It has been well-known that gravity and its supersymmetric extensions in 3-dimensions can be described by a Chern-Simons theories 
\cite{Deser:1981wh,Deser:1982sw,Witten:1988hc, Achucarro:1987vz}. 
Black hole in  such gravity theories have also been studied \cite{Izquierdo:1994jz}. Higher spin gravity with spins $2,3,\cdots N$ is described by generalizing the gauge group
from $SL(2,R)\times SL(2,R)$  to $SL(N,R)\times SL(N,R)$ \cite{Blencowe:1988gj,Vasiliev:2003ev,Bekaert:2005vh}. This can also be phrased in terms of the higher spin algebra $hs[\lambda]$ with $\lambda=-N$. These $hs[\lambda]$ theories have gained considerable interest in the context of its interpretation in terms of holographic minimal models \cite{Gaberdiel:2012uj}. As in the case for pure gravity one may also look for supersymmetric generalizations of higher spin gravity. The gauge group in such a case is given by $shs[\lambda]$ \cite{Creutzig:2011fe}. For the case of $\mathcal{N}=2$ higher spin supergravity the gauge group is given by $sl(N|N-1)$. The Chern-Simons action for such a supergravity theory based on a supergroup $\mathcal{G}$ is given by
 \begin{equation}
S=\frac{k}{2\pi} \int \left[ \text{str} \left(A\; dA +\frac{2}{3} A^3 \right)  -  \text{str} \left(\bar A\; d\bar A +\frac{2}{3}\bar  A^3 \right)    \right]
\end{equation}
where $A=A^a_\mu T^a dx^\mu$, with $T^a$ being the generators of the supergroup. The gauge connections are given in terms of the tetrad and vielbein of the background metric as
\begin{equation}
A=\omega + e \qquad \bar{A} = \omega -e
\end{equation}
 The equation of motion for the action above is the flatness conditions on the gauge connections and is given by
\begin{equation}
dA + A\wedge A =0 \qquad d\bar{A} + \bar{A} \wedge \bar{A} =0
\end{equation} 
The metric can be obtained from the gauge connection using 
\begin{equation}
g_{\mu\nu} = \frac{1}{\epsilon_{(N|N-1)}}\text{str}(e_\mu e_\nu)
\end{equation}
Here $\epsilon_{(N|N-1)}$ is a normalization constant given by $\frac{N(N-1)}{4}$.

\subsection{On the superalgebra $sl(3|2)$}
We shall be considering black holes in the simplest higher spin supergravity theory. This is based on the algebra $sl(3|2)$ and has $\mathcal{N}=2$ supersymmetry. This is the semiclassical version ($c\rightarrow \infty$) of the global part of the $\mathcal{N}=2$ super-$\mathcal{W}_3$ algebra \cite{Romans:1991wi}. This is a refection of the fact that the asymptotic algebra for this case is that of super-$\mathcal{W}_3$ and the dual superconformal field theory has this symmetry.

This algebra has bosonic generators $J$, $L$, $V$ and $W$ corresponding to spins 1, 2, 2 and 3 respectively. The algebra formed by just $L$ is the usual  $SL(2,R)$ algebra. There are fermionic generators $G^\pm$ and $U^\pm$ corresponding to spins 3/2 and 5/2 respectively. 

We shall list explicit commutation relations for this algebra below. 
\begin{eqnarray}\label{bos}
[J,J]&=&0, \qquad 
[L_m,L_n]=(m-n)L_{m+n},   \\ \nonumber
[V_m,V_n]&=& (m-n)(L_{m+n} + \k V_{m+n}), \\ \nonumber
[W_m,W_n]&=&\tfrac{1}{4} (m-n)(2m^2+2n^2-mn-8) ( L_{m+n} + \tfrac{\k}{5}  V_{m+n} ),  
\\ \nonumber 
 [ J, L_n ]&=&0,  \qquad
[J,V_n]=0,  \qquad 
[J, W_n]=0 ,   \\ \nonumber
[L_m,V_n]&=&(m-n)V_{m+n},  \qquad 
[L_m,W_n]=(2m-n)W_{m+n},  \\ \nonumber
[V_m,W_n] &=&\tfrac{\k}{5}(2m-n)W_{m+n}.  
\end{eqnarray}
where $\k=\pm (5/2)i$. Here the subscripts $m,n$ on the generators $L$ run from $-1, 0, 1$ while the 
subscripts on the generators $W$ run from $-2, -1, 0, 1, 2$. 
The commutation relations between bosonic and fermionic generators are given by 
\begin{eqnarray} \label{bosf}
[L_m,\pmG_r]&=&  (\tfrac{1}{2}m - r) \pmG_{m+n},  \qquad 
[J, G^\pm_r] = \pm G^\pm _{r},   \\ \nonumber
[L_m , U^\pm _r] &=&  (\tfrac{3}{2}m-r)U^\pm_{m+r},   \qquad
[J , U^\pm _r] = \pm U^\pm_{r},   \\ \nonumber
[V_m ,G^\pm_r ] &=& \pm U^\pm_{r+m},  \qquad
[G^\pm_r, W_m]=(2r -\tfrac{1}{2}m)U^\pm_{r+m},  \\ \nonumber
[V_m, U^+_r] &=& \tfrac{2}{5}\k ( \tfrac{3}{2} m -r ) \pU_{m+r} + \tfrac{1}{4} (3m^2 - 2mr + r^2 -\tfrac{9}{4} ) \pG_{m+r},  \\ \nonumber
[V_m, U^-_r] &=& -\tfrac{2}{5}\k^* ( \tfrac{3}{2} m -r ) \mU_{m+r} - \tfrac{1}{4} (3m^2 - 2mr + r^2 -\tfrac{9}{4} ) \mG_{m+r},  \\ \nonumber
[U^+_r,W_m]&=& \tfrac{\k}{10} (2r^2-2rm+m^2-\tfrac{5}{2})U^+_{r+m}   \\  \nonumber
& & \qquad + \tfrac{1}{8} ( 4r^3-3r^2 m +2rm^2-m^3 - 9r +\tfrac{19}{4}m ) G^+ _{r+m}, 
 \\ \nonumber
[U^-_r,W_m]&=& \tfrac{\k^*}{10} (2r^2-2rm+m^2-\tfrac{5}{2})U^-_{r+m} \\ \nonumber
& & \qquad + \tfrac{1}{8} ( 4r^3-3r^2 m +2rm^2-m^3 - 9r +\tfrac{19}{4}m ) G^- _{r+m}. 
\end{eqnarray}
Here the subscripts $r, s$ on $G^\pm$ run from $-1/, 1/2$ while the subscripts
on the generators $U^\pm$ run from $-3/2, -1/2, 1/2, 3/2$. 
Finally the anti-commutation rules between the fermionic generators are given by 
\begin{eqnarray}
\{  G^\pm_r, G^\mp_s \} &=&  2L_{r+s} \pm (r-s) J, \qquad  
\{  G^\pm_r , G^\pm_s  \} =0,  \\ \nonumber
\{   G^\pm_r , U^\mp _s      \}  &=&  2W_{r+s} \pm (3r-s)V_{r+s}, \qquad 
\{ G^\pm _r , U_s^\pm \} = 0 , \\ \nonumber
\{ \pU_r , \mU_s \} &=&  -\tfrac{2}{5}\k (r-s) W_{r+s}+( 3s^2-4rs+3r^2-\tfrac{9}{2} )( \tfrac{1}{2}L_{r+s}+\tfrac{\k}{5} V_{r+s}   )  \\ \nonumber
& & \qquad \ + \tfrac{1}{4} (r-s) (  r^2 +s^2 -\tfrac{5}{2} ) J_{r+s},  \\ \nonumber
\{  \pmU_r , \pmU_s  \} &=& 0 . 
\end{eqnarray}
This algebra has 12 fermionic and bosonic generators each. The bosonic part of the superalgebra is given a direct sum of the subalgebras $sl(3) \oplus sl(2) \oplus u(1)$. This can be explicitly seen by defining new spin-2 generators, $T^\pm$ in terms of $L$ and $V$ as follows
\begin{equation}\label{decouple-1}
T^+_m = -\frac{1}{3}(L_m + 2i V_m) \, \qquad T^-_m = \frac{1}{3}(4 L_	m + 2i V_m). 
\end{equation}
Substitituting these redefintions in (\ref{bos}) we  obtain
\begin{equation}
[T^+_m,T^-_n]=0 \ , \qquad [T^+_m,W_n]=0 , 
\end{equation}
The generators  $T^+_m$ obey the $sl(2)$ algebra while the 
generators $ T^-_n, W_m$ obey the  commutation relations of the $sl(3)$ algebra 
\begin{eqnarray}
 [T^-_m , T^-_n] &=& ( m-n) T^-_{m+n}, \qquad 
[T^-_m, W_n] = ( 2m -n) W_{m+n}, \\ \nonumber
[W_m , W_n] &=& \frac{3}{16} ( m-n) ( 2m^2 + 2n^2 -mn-8 ) T_{m+n}^-.
\end{eqnarray}
The $sl(3)$ algebra above is same as that of as the one  given in equation (A.2) of 
\cite{Gutperle:2011kf}  with  $\sigma = (3/4)^2$. For the $sl(3)$ part we shall be using the same representation of the generators as given in \cite{Gutperle:2011kf} while the for the $sl(2)$ part the representation in terms of Pauli matrices are used. For the $u(1)$ generator the diagonal matrix $(-2,-2,-2,-3,-3)$ is used as mentioned in Section 61 of \cite{Frappat:1996pb}. 
We will choose the gravitational $sl(2)$ generators to be that given by $L_m$ and the corresponding
supercharges to be $G^\pm_r$ which  form the principal embedding of  $osp(2|1)$ in  $sl(3|2)$. 
Throughout this paper we will work with this embedding, recently there has been 
a study of the other embeddings  of 
$osp(2|1)$ in $sl(3|2)$ in \cite{Peng:2012ae} and the renormalization group flows between the 
various embeddings. 

\subsubsection*{The decomposition of the super-$\cW_3$ algebra}

The fact that the global generators of the  bosonic part of 
super-$\cW_3$  decomposes into  global $sl(3)\oplus sl(2)\oplus u(1)$ 
also carries over to  local generators. 
It is known that  the 
bosonic subalgebra of the $\mathcal{N}=2$ super-$\cW_3$ algebra also splits into three mutually commuting pieces as mentioned in \cite{Romans:1991wi}. The splitting can be written as
\begin{align} \label{decoup1}
\mathcal{N}=2 \text{ super-}\cW_3 \supset  \cW^+ \oplus \cW^- \oplus u(1)
\end{align}
$\cW^+$ contains just a spin-2 generator which is 
given by  ($T^+ + \tfrac{1}{2c}J^2 $) in the 
large $c$ limit.   $\cW^-$ contains a spin-2 generator given by 
($T^- - \tfrac{2}{c}J^2 $) and a spin-3 generator   which is given by 
($ W - \tfrac{6}{c}JV $) in the large $c$ limit. 
 The central charges for these subalgebras are 
\begin{equation}
c^+=-\frac{1}{3}c \ , \quad c^-=\frac{4}{3}c \ , \quad c_J=1 
\end{equation}
We have written the field redefinitions and the central charges in the large $c$ limit. 
The full quantum version of this  decomposition  and the central charges in 
each of the commuting sectors are given in 
\cite{Romans:1991wi}. 
The \( \cW^+ \) is then a Virasoro algebra with central charge \( c^+\) while \(\cW^-\) is a \(\cW_3\) algebra with \(c^-\) as its central charge. This decomposition will
 play an important role when we evaluate 
the partition function of the black hole from the  dual CFT and show that it precisely agrees with 
that obtained from the classical solution in the bulk.

\subsection{Black holes in higher spin supergravity}

There has been many recent constructions of classical solutions (black holes and conical defects) in higher spin theories \cite{Gutperle:2011kf,Ammon:2011nk, Castro:2011iw}. Similar constructions in higher spin supergravity  were studied in \cite{Tan:2012xi,Datta:2012km}. They turn out to be interesting in their own right \cite{Castro:2011fm,Banados:2012ue,Kraus:2013esi,Ammon:2012wc}  and also from their interpretation in the dual CFT \cite{Kraus:2011ds, Gaberdiel:2012yb, Peng:2012ae}. The thermodynamics of higher spin black holes turn out to be quite interesting and have been recently investigated in \cite{Perez:2012cf,  Campoleoni:2012hp, David:2012iu, Chen:2012ba,  Perez:2013xi, deBoer:2013gz}.
Here we shall be interested in constructing black hole solutions and studying their thermodynamics in such higher spin supergravity theories. 
 The trial  gauge connections for the higher spin black hole embedded in $sl(3|2)$ which we shall be considering here are as follows
\begin{equation}
A=b^{-1} a(x^+) b + b^{-1} d b  \qquad \bar{A}=b^{-1} \bar{a}(x^+) b + b^{-1} d b 
\end{equation} 
 where 
\begin{align}
&a =  \left( L_1 - \frac{2\pi}{k} \cL L_{-1}  - \frac{2\pi}{k} \cV V_{-1}  + \frac{\pi}{2k\sigma} \cW W_{-2}  + {\frac{\pi}{k}} \cJ J_0  \right) dx^+ \nn \\
&\qquad +  \left( \mu W_2 + w_1 W_1 +w_0 W_0 + w_{-1} W_{-1} + w_{-2} W_{-2} +  \rho V_1 +v_0 V_0 + v_{-1} V_{-1} \right. \nn \\
&\qquad\qquad \left. + \ell L_{-1}- \gamma J_0  \right) dx^- \label{conn-a}\\
&\bar a =  - \left( L_1 - \frac{2\pi}{k} \bar \cL L_{-1}  - \frac{2\pi}{k}\bar \cV V_{-1}  + \frac{\pi}{2k\sigma} \bar\cW W_{-2}  -  {\frac{\pi}{k}} \cJ J_0    \right) dx^- \nn \\
& -  \left(\bar \mu W_{-2} +\bar  w_1 W_1 +\bar  w_0 W_0 + \bar  w_{-1} W_{-1} +\bar   w_{2} W_{2} + \bar   \rho V_1 +\bar   v_0 V_0 +\bar   v_{-1} V_{-1} \right. \nn \\
&\qquad\qquad \left.  +\bar   \ell L_{1}   + \gamma J_0  \right) dx^+
\end{align}
One is motivated to choose deformations of asymptotically-AdS connections of this form on basis of \cite{Campoleoni:2010zq, Gutperle:2011kf}. The 14 undetermined functions $$\{ \cJ,\cL,\cV,\cW,\mu,\ell,v_{-1},v_0,\rho,w_{-2},w_{-1},w_0,w_1,w_2 \}$$ are allowed to depend on $x^+(=t+\phi)$ and $x^-(=t-\phi)$. It will be seen later that $\gamma, \ \rho$ and $\mu$ are chemical potentials conjugate to $\cJ$, $\cV$ and $\cW$ \footnote{$\rho$ used in the connection $a$ is the chemical potential conjugate to $\cV$ and should not be confused with the radial coordinate. }.

\section{Bulk equations of motion and Ward identities}

In this section we shall be taking a look at the constraints on $\cJ$, $\cL$, $\cV$ and $\cW$ imposed by the flatness conditions. We then  show that we  will  obtain the same constraints using the Ward identities from the operator product expansions of the $\mathcal{N}=2$ super-$\cW_3$ algebra. In the process we will  relate the  charges and chemical potentials in the bulk to their counterparts in the boundary.  This method was  developed in bosonic theories in \cite{Gutperle:2011kf}
and applied to supersymmetric theories in \cite{Moradi:2012xd, Creutzig:2012xb}.
  The AdS/CFT dictionary for this case will be obtained.   
When these charges and currents are used as thermodynamic variables to calculate the entropy
and the partition function, 
the identifications of these quantities with that of the super-$\cW_3$ CFT 
will be important   to match the entropy and the partition function obtain from field theory calculations. 

\subsection{Bulk equations of motion}
On substituting the connection (\ref{conn-a}) in the Chern-Simons equation of motion ($ da + a \wedge a =0$) we get the following considering the coefficients of each of the generators 
\begin{align} \label{EOM}
w_1 &= - \pd_+ \mu \nn \\
w_0 &= +\frac{1}{2} \pd_+^2 \mu - \frac{4\pi}{k} (\cL + \tfrac{i}{2}\cV) \mu \nn \\
w_{-1} &= - \frac{1}{6} \pd ^3 \mu + \frac{4\pi}{3k} \pd_+ (\cL + \tfrac{i}{2}\cV) \mu + \frac{10\pi}{3k}(\cL + \tfrac{i}{2}\cV) \pd _+ \mu  \nn \\
w_{-2} &= \frac{1}{24} \pd^4_+ \mu - \frac{4\pi}{3k} \pd ^2_+  \mu (\cL + \tfrac{i}{2}\cV) -\frac{7\pi}{6k} \pd_+ (\cL + \tfrac{i}{2}\cV) \pd_+ \mu - \frac{\pi}{3k} \pd^2_+ (\cL + \tfrac{i}{2}\cV) \mu  \nn \\ &\qquad + \frac{4\pi ^2}{k^2}(\cL + \tfrac{i}{2}\cV)^2 \mu + \frac{i\pi \cW \rho}{4k\sigma} \nn \\
v_0 &= -\pd_+ \rho \nn \\
v_{-1} &= \frac{1}{2} \pd_+ ^2 \rho + \frac{3\pi i }{2k\sigma} \cW \mu - \frac{2\pi}{k}(\cL + \tfrac{5i}{2} \cV) \rho \nn \\
\ell &= \frac{3\pi}{k\sigma} \cW \mu - \frac{2\pi}{k} \cV \rho 
\end{align}
$\cL$, $\cV$ and $\cW$ satisfy the following equations 
\begin{align}
\pd_- \cJ &= \frac{k}{\pi}\pd_+ \gamma  \label{eom-j} \\
\pd _- \cL &= -\frac{8}{3}\pd _+ \cW \mu - 4 \cW \pd_+ \mu + \pd_+ \cV \rho + 2\cV \pd_+ \rho\label{eom-l} \\
\pd_- \cV &= -\frac{k}{4\pi} \pd^3_+ \rho - \frac{3i}{4\sigma} \pd _+ \cW \mu - \frac{9i}{8\sigma} \cW \pd_+ \mu + \pd_+(\cL + \tfrac{5i}{2}\cV)\rho + 2(\cL+ \tfrac{5i}{2}\cV)\pd_+ \rho \label{eom-v} \\
\pd_+ \cW &= \frac{3i}{2}\cW \pd_+ \rho + \frac{i}{2}\pd_+ \cW \rho + \frac{\sigma k}{12\pi}\pd_+^5 \mu - \frac{2\sigma}{3} \pd_+^3 (\cL + \tfrac{\kappa}{5}\cV)\mu  - 3\sigma \pd_+^2  (\cL + \tfrac{\kappa}{5}\cV) \pd_+ \mu \nn \\
& \quad - 5\sigma \pd _+ (\cL + \tfrac{\kappa}{5}\cV) \pd_+^2 \mu  -\frac{10\sigma}{3} (\cL + \tfrac{\kappa}{5}\cV)\pd_+^3 \mu + \frac{64\pi \sigma}{3k} (\cL + \tfrac{\kappa}{5}\cV)\pd_+ (\cL + \tfrac{\kappa}{5}\cV) \mu \nn \\ &\quad +  \frac{64\pi \sigma}{3k} (\cL + \tfrac{\kappa}{5}\cV)^2 \pd_+ \mu  \label{eom-w}
\end{align}

\subsection{Ward Identities from OPEs}
In this section we shall try to obtain the same equations as Ward identities from the OPEs of the $\mathcal{N}=2$ super-$\cW_3$ algebra \cite{Romans:1991wi}. The OPEs are listed in Appendix A.

\subsubsection*{Spin-1 current, $J$}
The non-vanishing OPEs for operators along with $J(z)$ are  that of $J(z)J(w)$ and $J(z)W(w)$. The Ward identity is given by 
\begin{equation}\label{j-ward}
 \pd_{\bz} J(z,\bz) = \pd_{\bz}  \int d^2 y \left(J(z)J(y) \gamma(y) + J(z) W(y) \mu(y) \right)
\end{equation}
here by $J(z,\bz)$ we mean the expectation value $\langle J \rangle_{\gamma,\rho,\mu}$, weighted with $e^{\frac{1}{2\pi}\int(\gamma J + \rho V + \mu W )}$. 
Upon using the OPEs  and upon using $\pd_\bz \left( \frac{1}{z}\right)=2\pi \delta^{(2)}(z,\bz)$ we get 
\begin{equation}
 \pd_{\bz} J = -  \frac{c}{3} \pd_z \lb  \gamma +\frac{6}{c} V\mu  \rb 
\end{equation}
On converting the above equation to the Lorentzian signature $
\pd_- \rightarrow -\pd_\bz \ , \pd_+ \rightarrow \pd_z 
$
and using $\frac{k}{2\pi}=\frac{c}{6}$  we can identify the bulk variables in (\ref{eom-j}) with those from the algebra as
\begin{equation}\label{id1}
\cJ = J \ ,  \qquad \gamma_{\text{bulk}}= \gamma + \frac{6}{c}V\mu \   .
\end{equation}

\subsubsection*{Stress-tensor, $T$}
For this case the non-zero OPEs are that of $T(z)J(y)$, $T(z)V(y)$ and $T(z)W(y)$. The variation is then given as
\begin{equation}
\pd_\bz T(z,\bz) = \pd_{\bz}  \int d^2 y \left(T(z)J(y) \gamma(y) + T(z) V(y) \rho(y) + T(z) W(y) \mu(y) \right)
\end{equation}
Substituting the OPE we obtain 
\begin{equation} \label{t-ward}
 \pd _\bz T = - ( 2\pd_z W \mu + 3 W \pd _z \mu + 2 V \pd _z \rho + \pd _z V \rho + J \pd _z \gamma)
\end{equation}
Using (\ref{j-ward}), converting to Lorentzian signature and comparing with (\ref{eom-l}) leads us to
\begin{align}\label{id2}
\cL =& -\left(  T - \frac{3}{2c}J^2  \right),  \quad  \cV=-V,  \quad  \cW=\frac{3}{4}\left( W - \frac{6}{c}JV \right),  \nn \\ 
&\qquad\quad \rho_{\text{bulk}}=\rho+\frac{6}{c}J\mu  , \qquad  \mu_{\text{bulk}}=\mu 
\end{align}
along with the ones in (\ref{id1}) which remain consistent identifications. Thus, the Ward identities of $J$ and $T$ are sufficient to give all the identifications between the bulk and algebraic variables. We shall now find  the Ward identities of $V$ and $W$ and compare them with the constraints from the bulk equations of motion with the identifications  given in (\ref{id1}) and (\ref{id2}). This 
is performed as an additional consistency check of the identifications. 

\subsubsection*{The other spin-2 current, $V$}
The contributing OPEs in this case are that of $V(z)V(y)$ and $V(z)W(y)$. 
\begin{equation}
\pd_\bz V(z,\bz) = \pd_{\bz}  \int d^2 y \left(V(z)V(y) \rho(y) + V(z) W(y) \mu(y) \right)
\end{equation}
from which we get
\begin{align}\label{v-ward}
 \pd _\bz V &= - \lb \frac{c}{12} \pd ^3 _z \rho + 2(T+\kappa V - \tfrac{3}{2c}J^2)\pd_z \rho + \rho \pd_z  (T+\kappa V - \tfrac{3}{2c}J^2)  \right. \nn \\
& \left. + \frac{2}{c}(J\pd_z(T+\kappa V)-2\pd_z J(T+\kappa V))\mu + 3\pd_z \mu C^{[3]} +2\pd_z  C^{[3]} \mu + \pd^3_z (\tfrac{1}{2}J\mu)      \rb
\end{align}	
where $C^{[3]}$ is given in equation (\ref{abc}) in the appendix. The identifications in (\ref{id1}) and (\ref{id2}) can be consistently used in while comparing (\ref{eom-v}) with the Lorentzian version of the above equation and using (\ref{j-ward}) and (\ref{t-ward}).
\def\cT{\mathcal{T}}
\subsubsection*{The spin-3 charge, $W$}
We finally come to the case of the spin-3 charge, $W$. The contributions to the variation arise from $W(z)W(y)$, $W(z)V(y)$ and $W(z)J(y)$ OPEs
\begin{equation}
\pd_\bz W(z,\bz) =  \pd_{\bz} \int d^2 y \left(W(z)W(y)\mu(y) \rho(y) + W(z) V(y) \rho(y) + W(z)J(y) \gamma(y) \right)
\end{equation}
Substituting the OPEs we get the following
\begin{align} \label{w-ward}
\pd_\bz W(z,\bz) =& - \frac{c_W}{360} \pd_z^5 \mu - 10 B^{[2]}\pd^3_z \mu - 15\pd_z B^{[2]} \pd_z^2 \mu - (2B^{[4]}+9\pd_z^2 B^{[2]})\pd_z \mu  \nn \\
& -( \pd_z B^{[4]} + \pd_z^3 B^{[2]} )\mu - C^{[1]} \pd_z^3 \rho - 3(C^{[3]} - 5\pd_z C^{[2]} )\pd_z \rho -(C^{[4]}+\pd_z C^{[3]})\rho \nn \\
& + 2V \pd_z \gamma 
\end{align}
Substituting the $B^{[\, ]}$s and $C^{[\, ]}$s above from the equation (\ref{abc}) and upon using (\ref{j-ward}), (\ref{t-ward}), (\ref{v-ward}) we can see that this equation exactly matches with (\ref{eom-w}) with the identifications given in (\ref{id1}) and (\ref{id2}). 
In doing this we require to make an additional identification $c_W=\frac{45}{2\pi}k$ . 
This is consistent with the relation   \(c_W = \tfrac{15}{2}c\) found in \cite{Romans:1991wi} 
which was arrived at using the Jacobi identities satisfied by the super-${\cal W}_3$ algebra.  
Although the calculations for this specific Ward identity  are quite involved, one can explicitly see the matching by comparing each of the coefficients of the derivatives of the chemical potentials $\gamma$, $\rho$ and $\mu$. 

\subsection{Summary -- The AdS/CFT dictionary }
The identifications of the charges and chemical potentials in the boundary and the bulk are summarized in the following table \vspace{.1cm}
\begin{center}
\begin{small}
\begin{tabular}{|c|c|c|c|c|}
\hline
&Spin & Bulk & Boundary \\ \hline\hline
&1 & $\cJ$ & $J$ \\  
Charges &2 & $\cL$ & $-(T-\tfrac{3}{2c}J^2)$ \\ 
&2 & $\cV$ & $-V$ \\  
&3 &$\cW$ & $\tfrac{3}{4}(W-\tfrac{6}{c}JV)$\\   \hline\hline
Chemical&1 &$\gamma$ &$\gamma+\tfrac{6}{c}V\mu$\\
potentials&2 &$\rho$ &$\rho + \tfrac{6}{c}J\mu$\\
&3 &$\mu$ &$\mu$\\ \hline 
\end{tabular}
\end{small}
\\ \vspace{0.3cm}
\begin{footnotesize}
\textbf{Table 1} : Relating the charges and currents of the bulk with that of the CFT\footnote{Note that in our conventions which is the same as in \cite{Gutperle:2011kf}, $-(T-\tfrac{3}{2c}J^2)$  is positive}
\end{footnotesize}
\end{center}
These identifications constitute the AdS/CFT dictionary for this higher spin black hole background. It is interesting to note that the charges and chemical potentials shift due to the presence of the R-symmetry part. 
Note that these 
identifications satisfied several non-trivial checks.  
The  structure of the ${\cal N}=2$ super-${\cal W}_3$  OPE's was used to 
obtain the identifications. The OPE's are of-course constrained by supersymmetry. 
This is an independent reason for the 
dual CFT to have the ${\cal N}=2$ super-${\cal W}_3$ symmetry. 

Motivated by the definition (\ref{decouple-1}) for the decoupled generators, let us now define the 
following linear combinations of the charges in the bulk. 
\begin{equation}\label{newcharge}
\cT_+=-\tfrac{1}{3} ( \cL+2 i \cV ) \ , \quad \cT_-= \tfrac{4}{3}  (\cL + \tfrac{i}{2} \cV) \ , \quad \cW_-=\tfrac{4}{3}\cW
\end{equation}
the identifications for the charges become
\begin{equation}
\cT_+ \rightarrow T_+ + \tfrac{1}{2c}J^2 \ , \quad \cT_- \rightarrow T_- - \tfrac{2}{c} J^2 \ , \quad \cW_- \rightarrow W-  \tfrac{6}{c} JV
\end{equation}
Now the bulk charges ${\cal T}_\pm, {\cal W}_{-} $
are identified  precisely with the combinations of the boundary currents for which 
 the bosonic part of the \(\mathcal{N}=2\) super-\(\cW_3\) algebra
 decoupled into 3 bosonic sub-algebras 
  as  mentioned  in subsection 2.2. 
  This shows the natural identification of the decoupled operators in the 
  CFT is in terms of the `decoupled charges' in the bulk. The combination of the boundary 
  currents which results in the three decoupled bosonic algebra can also be 
  thought of as cosetting out the $U(1)$  \footnote{This point was mentioned to us
  by Matthias Gaberdiel.}. The reason is that, now currents $T_+ + \tfrac{1}{2c}J^2 $ and
  $T_- -  \tfrac{2}{c} J^2$ and $ W-  \tfrac{6}{c} JV$ are uncharged with respect to the $U(1)$. 
  Let us again emphasize that this decoupling resulted due to the tight structure of the 
  \(\mathcal{N}=2\) super-\(\cW_3\) OPE's. 
  This phenomenon of decoupling is a property  of the \(\mathcal{N}=2\) super-\(\cW_3\)
  algebra and is present in the supersymmetric minimal models dual to Chern-Simons
  theories based on the infinite dimensional supergroup based on 
  $shs[\lambda]$ as discussed in section 5.4.

\section{The higher spin black hole}
\subsection{The gauge connections}
In this section we write down the connections of the higher spin black hole embedded in the $sl(3|2)$ theory. If all the charges and chemical potentials are assumed to be independent of $x_\pm$, then from the equations of motion (\ref{EOM}) we get the following solution
\begin{align}  \label{gc1}
a =&  \left( L_1 - \frac{2\pi}{k} \cL L_{-1}  - \frac{2\pi}{k} \cV V_{-1}  + \frac{\pi}{2k\sigma} \cW W_{-2}  + {\frac{\pi}{k}} \cJ J_0  \right) dx^+ \nn \\
&+  \left( \mu W_2 - \frac{4\pi}{k} (\cL +\tfrac{i}{2}\cV)\mu  W_0 + (\tfrac{4\pi^2}{k^2} (\cL+\tfrac{i}{2}\cV)^2 \mu + \tfrac{i\pi}{4k\sigma}\cW \rho)W_{-2} +  \rho V_1 \right. \nn \\
&\qquad  \left.  +(\tfrac{3\pi i}{2k\sigma}\cW\mu  -  \tfrac{2\pi}{k}(\cL + \tfrac{5i}{2}\cV)  \rho) V_{-1} + (\tfrac{3\pi}{k\sigma}  \cW\mu-\tfrac{2\pi}{k}\cV \rho) L_{-1}- {\frac{\pi}{k}} \cJ J_0  \right) dx^-   
\end{align}\vspace{-.5cm}
\begin{align}
\bar a =&  - \left( L_{-1} - \frac{2\pi}{k} \bar \cL L_{1}  - \frac{2\pi}{k}\bar \cV V_{1}  + \frac{\pi}{2k\sigma} \bar\cW W_{2}  -  {\frac{\pi}{k}} \cJ J_0    \right) dx^- \nn \\
&-  \left( \bar\mu W_{-2} - \frac{4\pi}{k} (\bar\cL +\tfrac{i}{2}\bar\cV)\mu  W_0  + (\tfrac{4\pi^2}{k^2} (\bar\cL+\tfrac{i}{2}\bar\cV)^2 \mu + \tfrac{i\pi}{4k\sigma}\bar\cW \bar \rho) W_{2} +  \bar \rho V_{-1} \right. \nn \\
&\qquad   +(\tfrac{3\pi i}{2k\sigma}\cW\mu  -   -\frac{2\pi}{k}(\bar\cL + \tfrac{5i}{2}\bar\cV) \bar \rho) V_{1}   \left. + (\tfrac{3\pi}{k\sigma} \bar\cW\bar\mu-\tfrac{2\pi}{k}\bar\cV\bar \rho) L_{1} + {\frac{\pi}{k}} \cJ J_0  \right)  dx^+
\end{align}
The above connection reduces to that of the charged BTZ black hole embedded in the gravitational $sl(2)$ for which the connections are given solely in terms of $L$ and $J$ generators. 
This can be seen by setting  $\mu=\bar{\mu}=\cW=\bar{\cW}=\rho= \bar \rho = {\cal V}= \bar{\cal V}=0$. Note that this black hole is however not continuously connected to the higher spin black hole of \cite{Gutperle:2011kf}.  The reason is that even though if one sets 
$\rho= \bar \rho = {\cal V}= \bar{\cal V}=0$, the coefficient of $V_{-1}$ does not vanish.

\def\tr{{\text{tr}}} 
\def\cT{{\mathcal{T}}}
\subsection{Black hole holonomy and integrability}
We shall now investigate the holonomy for the higher spin black hole constructed. The holonomy is a gauge invariant and meaningful quantity in Chern-Simons theory. We shall also see that this observable will play a  role in determining the entropy for the black hole. We shall be determining the holonomies along the thermal circle and then demand their eigenvalues to be the same as the one for the BTZ black hole. 

The holonomy around the Euclidean time circle $(z,\bz)\rightarrow(z+2\pi \tau , \bz + 2\pi\bar{\tau})$ is 
\begin{align}
H=b^{-1} e^\omega b \ , \qquad \bar H=b e^{\bar{\omega}} b^{-1}
\end{align}
where
\begin{equation}
\omega=2\pi(\tau a_+ - \bar\tau a_-) \ , \qquad \bar{\omega}=2\pi(\tau\bar{a}_+ - \bar{\tau}\bar{ a}_-)
\end{equation}
For the gauge connections embedded in the $sl(3|2)$ theory, $\omega$ has the block diagonal form $sl(3)\oplus sl(2)$. The eigenvalues of the holonomy matrix for the BTZ black hole is $(-2\pi i, 0, 2\pi i)$ and $(-\pi i, \pi i)$.

On defining 
\begin{equation}\label{chem1}
\alpha_1=\bar{\tau}\gamma, \quad \alpha_2=\bar{\tau}\rho, \quad \alpha_3=\bar{\tau}\mu 
\end{equation}
the $sl(2)$ part of the holonomy for the gauge connection given in (\ref{gc1}) is 
\begin{equation}
\omega_{sl(2)}=\left(
\begin{array}{cc}
 -\frac{6 \pi  (\alpha_1  k+\pi  \cJ \tau )}{k} & \frac{4 \pi ^2 (\cL+2 i \cV) (\tau -2 i \alpha_2 )}{k} \\
 2 \pi  (\tau -2 i \alpha_2 ) & -\frac{6 \pi  (\alpha_1  k+\pi  \cJ  \tau )}{k}
\end{array}
\right) \label{sl2-p}
\end{equation}
On finding the eigenvalues of the above matrix and equating them $(i\pi,-i\pi)$ we get\footnote{There exist two possibilities here regarding whether to equate an eigenvalue to $i\pi$ or $-i\pi$. However, the condition on $\cJ$ and $\alpha_1$ we write down is common to both the cases.}
\begin{equation}
\alpha_1  k+\pi  \cJ \tau =0  \label{u1-cond}
\end{equation}
Then (\ref{sl2-p}) becomes
\begin{equation}
\omega_{sl(2)}=\left(
\begin{array}{cc}
 0 & \frac{4 \pi ^2 \left(\tau - 2 i \alpha _2\right) (\cL+2 i \cV)}{k} \\
 2 \pi  \left(\tau - 2 i \alpha _2\right) & 0
\end{array}
\right)
\end{equation}
while the $sl(3)$ part is given by 
\begin{equation}
\omega_{sl(3)}=\left(
\begin{array}{ccc}
 \frac{16 \pi ^2 \left(\cL+\frac{i \cV}{2}\right) \sqrt{-\sigma } \alpha _3}{3 k} & \frac{\pi ^2 \left(8 \left(\cL+\frac{i \cV}{2}\right) \sigma  \left(\tau -\frac{i \alpha _2}{2} \right)+9 \cW \alpha _3\right)}{k \sigma } & \frac{4 \pi ^2 \left(16 \pi  \left(\cL+\frac{i \cV}{2}\right)^2 \sigma  \alpha _3-2 k \cW \left(\tau -\frac{i \alpha _2}{2} \right)\right)}{k^2 \sqrt{-\sigma }} \\
 2 \pi  \left(\tau -\frac{i \alpha _2}{2}\right) & -\frac{32 \pi ^2 \left(\cL+\frac{i \cV}{2}\right) \sqrt{-\sigma } \alpha _3}{3 k} & \frac{\pi ^2 \left(8 \left(\cL+\frac{i \cV}{2}\right) \sigma  \left(\tau -\frac{i \alpha _2}{2}\right)+9 \cW \alpha _3\right)}{k \sigma } \\
 -4 \pi  \sqrt{-\sigma } \alpha _3 & 2 \pi  \left(\tau -\frac{i \alpha _2}{2}\right) & \frac{16 \pi ^2 \left(\cL+\frac{i \cV}{2}\right) \sqrt{-\sigma } \alpha _3}{3 k}
\end{array}
\right)
\end{equation}
The combinations appearing above motivates us to define a new set of variables as follows
\begin{align} \label{l1l2}
\eta_+ = \tau - 2i\alpha_2 \ , \qquad &\cT_+=-\tfrac{1}{3} ( \cL+2 i \cV )   \ , \qquad k_+ =-\tfrac{1}{3}k    \nn \\
\eta_- =\tau - \tfrac{i}{2}\alpha_2 \ , \qquad &\cT_-= \tfrac{4}{3}  (\cL + \tfrac{i}{2} \cV) \ , \qquad \ \  k_- =\tfrac{4}{3}k  \\
&\cW_- = \tfrac{4}{3} \cW \nn
\end{align}
The combinations for $\cT_{\pm}$ and $\cW_-$ were mentioned previously in (\ref{newcharge}). This is due to the fact that the bosonic part of $sl(3|2)$ splits as $sl(3)\oplus sl(2) \oplus u(1)$.

We shall now demand that the eigenvalues of holonomy should equal that of the BTZ black hole. An equivalent way of saying this is 
\begin{align}
\tr(\omega^2_{sl(2)})&=-2\pi^2, \\  \quad \tr(\omega^2_{sl(3)})&=-8\pi^2, \\  \quad \det(\omega_{sl(3)})& =0 .
\end{align}
The first condition gives 
\begin{equation} \label{c-sl2}
\eta_+=\frac{ik_+}{2} \frac{1}{\sqrt{2\pi k_+ \cT_+}}
\end{equation}
the second condition for $\alpha_3=0=\cW_-$ gives
\begin{equation} \label{c-sl3}
\eta_- |_{{\alpha=0=\cW_-}}=\frac{ik_-}{2} \frac{1}{\sqrt{2\pi k_- \cT_-}}
\end{equation}
for non-zero $\alpha_3$ and $\cW$ we have
\begin{equation}\label{cond-trace}
256\pi^2 \sigma \alpha_3^2 \cT_-^2 - 24\pi k_- \eta_-^2 \cT_- -72\pi k_- \eta_- \alpha_3 \cW_- - 3k_-^2 =0
\end{equation}
and from the determinant condition we get 
\begin{align}
2048\pi^2 \sigma^2 \alpha_3^3 \cT_-^3 + 576\pi\sigma k _- \eta_-^2 \alpha_3 \cT_-^2  &+ 864\pi \sigma k_- \alpha_3^2 \eta_- \cW_- \cT_- \nn \\
& +864\pi\sigma k_- \alpha_3^3 \cW_-^2 -27k_-^2 \eta_-^3 \cW_- =0
\end{align}
These are the same conditions as (5.14) of \cite{Gutperle:2011kf} with $\cL\rightarrow\cT_-$,  $\tau \rightarrow\eta_-$ and $\alpha\rightarrow\alpha_3$. The integrability conditions are then given by
\begin{equation}\label{inte}
\frac{\pd \cW_-}{\pd \eta_-} = \frac{\pd\cT_-}{\pd\alpha_3}
\end{equation}

\subsection{Supersymmetry of the higher spin black hole}

We now analyse the supersymmetry of this higher spin black hole. In \cite{Datta:2012km} it was shown that the Killing spinor can be written in terms of products of background holonomies and odd-roots of the superalgebra. For calculational simplicity we shall restrict ourselves to the case $\cW=0$.

The matrix $a_\phi$ is given in terms of the Cartan matrices $H_i$   of the superalgebra \cite{Frappat:1996pb} as 
\begin{align}
S a_\phi S^{-1} = & \left( - \frac{4\pi i \mu \cT_-}{k} + i(\rho + 2i) \sqrt{\frac{2\pi\cT_-}{k}} \right) H_1 +  \left(  \frac{4\pi i \mu \cT_-}{k} + i(\rho + 2i) \sqrt{\frac{2\pi\cT_-}{k}} \right) H_2 \nn \\
& \qquad + (2\rho + i) \frac{2\pi\cT_+}{k}H_{\bar 4} + \lb  \frac{\pi \cJ}{k} +\gamma  \rb J
\end{align}
where $\cL_{1,2}$ are defined in (\ref{l1l2}). Using the subalgebra for $sl(3|2)$ one can derive the supersymmetric conditions using 
\begin{equation}\label{odd-root}
\lambda^r \alpha^r_i \pm \lb  \frac{\pi \cJ}{k} +\gamma  \rb = in_i \qquad n_i \in \mathbb{Z}
\end{equation}
where $r$ is the index for the Cartan matrices and $i$ is the fermionic direction which one chooses. $\alpha^r_i$s are the odd-roots of the superalgebra.
\def\bi{{\bar{\iota}}}
The commutation relations with the Cartan matrices with the fermionic generators are
\begin{align}
[H_1, E_{\bi,1}]&=E_{\bi,1}, \quad [H_1, E_{\bi,2}]=-E_{\bi,2}, \quad [H_1, E_{\bi,3}]=0 \nn \\
[H_2, E_{\bi,1}]&=0, \quad [H_2, E_{\bi,2}]=E_{\bi,2}, \quad [H_3, E_{\bi,3}]=-  E_{\bi,3} \\
&[H_{\bar{4}},E_{\bar{4}i}]=-E_{\bar{4}i}, \quad [H_{\bar{4}},E_{\bar{5}i}]=-E_{\bar{5}i} \nn
\end{align}
The supersymmetric condition(s) for the fermionic direction $\bi=\bar{4},\ j=2$ using (\ref{odd-root}) then turn out to be 
\begin{align}
- \left( - \frac{4\pi i \mu \cT_-}{k} + i(\rho + 2i) \sqrt{\frac{2\pi\cT_-}{k}} \right)   + & \left(  \frac{4\pi i \mu \cT_-}{k} + i(\rho + 2i) \sqrt{\frac{2\pi\cT_-}{k}} \right) \nn \\
&- (2\rho + i) \frac{2\pi\cT_+}{k}  + \lb  \frac{\pi \cJ}{k} +\gamma  \rb = i n
\end{align}
One can similarly find 5 more conditions for the other fermionic directions.

\def\siii{{sl(3)}}
\def\sii{{sl(2)}}
\section{Black hole thermodynamics}
\subsection{Entropy in terms of higher spin charges}
The integrability conditions (\ref{inte}) can be used as Maxwell relations for studying thermodynamics of black holes in higher spin supergravity. We are thus led to define the conjugate variable relations as follows
\begin{equation}\label{tau2}
\eta_- = \frac{i}{4\pi^2} \frac{\pd S_{\siii}}{\pd \cT_-} \ , \quad \alpha_3 =  \frac{i}{4\pi^2} \frac{\pd S_{\siii}}{\pd \cW_-}
\end{equation}
and also 
\begin{equation} \label{tau1}
\eta_+ = \frac{i}{4\pi^2} \frac{\pd S_\sii}{\pd \cT_+}
\end{equation}
The total entropy from the connection $A$ is given by
\begin{equation}
S= S_\sii(\cT_+) +S_\siii(\cT_-,\cW_-)
\end{equation}
This can be understood as follows. The gauge connection we are considering has a block diagonal form. The Chern-Simons action can then be written as a sum of two contributions from each of these blocks. Since the partition function involves $e^{-I}$, the entropy can be written as the sum above.

Using (\ref{c-sl2}) in (\ref{tau1}) we get
\begin{equation}
S_\sii = 2\pi \sqrt{2\pi k_+ \cT_+}
\end{equation}
The entropy contribution from the $sl(3)$ part can calculated as follows. We substitute (\ref{tau2}) in the trace condition (\ref{cond-trace}) and then solve the resulting differential equation. On basis of (\ref{c-sl3})  and \cite{Gutperle:2011kf} one can choose the following anzatz 
\begin{equation}
S_\siii = 2\pi \sqrt{2\pi k_- \cT_-} f(y) \qquad \text{where, }y=-\frac{27k_-\cW_-^2}{64\sigma \pi \cT_-^3}
\end{equation}
Substituting this in the differential equation we get
\begin{equation}
36y(2-y) (f')^2 + f^2 -1 =0
\end{equation}
which has the solution
\begin{equation}
f(y)=\cos \left(  \frac{1}{6} \arctan \left(  \frac{\sqrt{y(2-y)}}{1-y}  \right)   \right)
\end{equation}

Thus the final result for the entropy is 
\begin{align}\label{en-gravity}
S =& 2\pi\left( \sqrt{2\pi k_+ \cT_+} +  \sqrt{2\pi k_- \cT_-}\; f\left(-\frac{27k_-\cW_-^2}{64\sigma \pi\cT_-^3}\right) \right.\nn \\
& \quad \quad\left. + \sqrt{2\pi k_+ \bar\cT_+} +  \sqrt{2\pi k_-   \bar\cT_-} \;f\left(-\frac{27k_- \bar\cW_-^2}{64\sigma \pi  \bar\cT_-^3}\right)  \right)
\end{align}
This has the series expansion
\begin{align}\label{series-entropy}
S=& 2\pi \sqrt{2  \pi  k_+\cT_+} + 2\pi \sqrt{2  \pi  k_-\cT_-} \left(1+\frac{3 k_- }{256 \pi\sigma }\frac{\cW_-^2}{ \cT_-^3 } -\frac{105 k_-^2  }{131072 \pi ^2 \sigma ^2}  \frac{\cW_-^4}{ \cT_-^6}   + \cdots \right) \nn \\&\quad + \text{barred part}
\end{align}
Note that setting ${\cal W}_- = {\cal V} =0$, the entropy reduces to 
\begin{equation}
S= 2\pi\sqrt{ 2\pi k {\cal L }}.
\end{equation}
Here we have used the definitions in (\ref{l1l2}) to 
rewrite the expression in terms of $k$ and ${\cal L }$. 
Thus we have  obtained the expected answer of the entropy of the BTZ black hole embedded in the 
gravitational $sl(2)$ (constructed out of $L_m(=T^+_m + T^-_m)$ generators). This serves as an additional check of the fact  that the 
 the entropy is the sum of the contributions from the 
$sl(2)$ and the $sl(3)$ part of the Chern-Simons action.  

Recently there has been a discussion of entropy of higher spin black holes 
using the canonical definition of energy
in \cite{Perez:2013xi,deBoer:2013gz} called the `canonical formalism'. 
This method differs from that used by 
\cite{Gutperle:2011kf,Ammon:2011nk} which relies on the existence of the 
partition function and the compatibility with the first law called the `holomorphic formalism'. 
The difference arises due to choice of boundary terms. 
As mentioned in \cite{deBoer:2013gz}, the method \cite{Gutperle:2011kf,Ammon:2011nk} 
is more suited from the CFT point of view
due to the existence of the partition function. The CFT computations
agree precisely with that of the holomorphic formalism \cite{Gaberdiel:2012yb}. 
Our goal in this paper is also to explain the bulk partition function using the CFT. 
Therefore we have generalized the holomorphic formalism of  \cite{Gutperle:2011kf} to obtain 
the entropy of the supersymmetric black hole.

\def\gr{{\text{gravity}}}
\subsection{Entropy and partition function in terms of chemical potentials}
We wish to write the formula we just obtained for the entropy of the black hole (\ref{en-gravity}) in terms of the chemical potentials. This shall enable us to match the answer from conformal field theory. The partition function of the higher spin black hole embedded in $sl(3|2)\oplus sl(3|2)$ is\footnote{We are considering just one $sl(3|2)$ or the holomorphic part. }
\begin{equation}
Z_\gr = \Big{\langle}  e ^{2\pi i (\alpha_1 \cJ + \eta_+ \cT_+ + \eta_- \cT_- + \alpha_3 \cW_-)}  \Big{\rangle}_{\text{BH}}
\end{equation}
This can be written in terms of the conventional charges and chemical potentials as
\begin{equation}
Z_\gr =\Big{\langle} e ^{2\pi i   \lb \tau \cL + \alpha_1 \cJ - \alpha_2 \cV + \tfrac{4}{3}\alpha_3 \cW \rb }  \Big{\rangle}_{\text{BH}}
\end{equation}
here `BH' indicates that the quantity is evaluated for the higher spin black hole background. We thus have the following conjugate variable relations
\begin{align}
 \alpha_1 = \frac{i}{4\pi^2} \frac{\pd S}{\pd \cJ}\ &, \quad  \frac{\pd \log Z}{\pd \alpha_1}=4 \pi^2 i \cJ \\
 \eta_+ = \frac{i}{4\pi^2} \frac{\pd S}{\pd \cT_+}\ &, \quad  \frac{\pd \log Z}{\pd \eta_+}=4 \pi^2 i \cT_+ \\
 \eta_- = \frac{i}{4\pi^2} \frac{\pd S}{\pd \cT_-}\ &, \quad  \frac{\pd \log Z}{\pd \eta_-}=4 \pi^2 i \cT_- \\
 \alpha_3 = \frac{i}{4\pi^2} \frac{\pd S}{\pd \cW}\ &, \quad  \frac{\pd \log Z}{\pd \alpha_3}=4 \pi^2 i \cW 
\end{align}

In \cite{Kraus:2011ds} it was shown that the partition function for the hs$[\lambda]$ black hole could be written in terms of the chemical potentials. The approach taken was to assume a power series solution for $\cL$, $\cW$ and other higher spin charges, plug them  into the holonomy conditions and solve for the unknown coefficients in the power series expansions. Finally upon integrating the solutions for the charges, one can obtain the partition function. 
On performing a similar analysis as the above we obtain (for   the holomorphic part with terms up to $O(\alpha_3^4)$)
\begin{align}
\cT_+ &=-\frac{c^+}{48 \pi \eta_+ ^2} \label{ctp}	\\
\cT_- &=-\frac{c^-}{48 \pi \eta_- ^2}		-\frac{5 c^- \sigma }{36 \pi \eta_-^6} \alpha_3^2	-\frac{10 c^- \sigma ^2}{9 \pi \eta_- ^{10}} \alpha_3^4 	+ \cdots	\label{ctm}\\
\cW_- &=\frac{  c^- \sigma }{18 \pi \eta_- ^5}\alpha_3 +  \frac{40c^- \sigma ^2}{81 \pi  \eta_- ^9}  \alpha_3  ^3 + \cdots \label{cw}
\end{align}
The holonomy condition imposed the following relation on $\cJ$ and $\alpha_1$.
\begin{equation}
\alpha_1 k + \pi \cJ \tau =0 
\end{equation}
from which we get
\begin{equation}\label{two}
\alpha_1= - \frac{3\tau \cJ}{c} \ , \quad \cJ = - \frac{c \alpha_1}{3\tau}
\end{equation}
To study the thermodynamics, we have the following relations
\begin{equation}\label{three}
\tau  = \frac{i}{4\pi^2} \left(   \frac{\pd S}{\pd \cL} \right)_{\cL,\cV,\cW}\ , \quad  \alpha_1 = \frac{i}{4\pi^2}\left( \frac{\pd S}{\pd \cJ}  \right)_{\cJ,\cV,\cW}.
\end{equation}
Using it in the second equation of (\ref{two}) we get
\begin{equation}\label{four}
\cJ = - \frac{c}{3} \left(   \frac{\pd S}{\pd \cJ} \right)_{\cL,\cV,\cW} \left( \frac{\pd S}{\pd \cL}  \right)_{\cJ,\cV,\cW} ^{-1} {=}\ \frac{c}{3} \lb \frac{\pd \cL}{\pd \cJ} \rb _{S,\cV,\cW} \ {=} \ 0 
\end{equation}
This implies that the $u(1)$ charge is { forced to be zero} by the holonomy condition and it does not contribute to the partition function.  This conclusion can also be reached from the second relation 
in (\ref{three}). Since, the entropy is independent of $\cal J$ we obtain  $\alpha_1=0$. 
From the holonomy conditions this then implies ${\cal J}=0$. 
However it was important to keep track of the ${\cal J}$ to obtain the redefinitions
given in Table 1. 

On integrating (\ref{ctp}), (\ref{ctm}) and (\ref{cw}) we obtain the partition function to be
\begin{align}
\log Z_\gr (\eta_+,\eta_-,\alpha_3) &= \frac{\pi i c^+}{12\eta_+} + \frac{\pi i c^-}{12 \eta_-} \lb 1+ \frac{4\sigma}{3} \frac{\alpha_3^2}{\eta_-^4}+ \frac{160\sigma^2}{27}\frac{\alpha_3^4}{\eta_-^8} + \cdots \rb 
\end{align}
where we have used $k_\pm = c_\pm /6$. On exponentiating the above, we get
\begin{align}\label{Z-bh}
Z_\gr = q^{-c/24} \lb 1+ \frac{2\pi i c^-}{\eta_-} \lb  \frac{\sigma}{18}\frac{\alpha_3^2}{\eta_-^4} - \frac{20\sigma^2}{81}\frac{\alpha_3^4}{\eta_-^8}+\cdots \rb  \rb
\end{align}
Upon using $S= \log Z (\eta_+,\eta_-,\alpha_3) - 4\pi^2 i(\eta_+ \cT_+ + \eta_- \cT_- + \alpha_3 \cW)$, the entropy is
\begin{align}\label{en-chem}
S (\eta_+,\eta_-,\alpha_3) =\frac{i \pi  c^+}{6 \eta_+ } + \frac{i \pi  c^-}{6 \eta_- } \lb 1+ \frac{8\sigma}{3} \frac{\alpha_3^2}{\eta_-^4}+ \frac{160\sigma^2}{9}\frac{\alpha_3^4}{\eta_-^4} + \cdots \rb 
\end{align}
The equivalence of the above formula with the one in terms of the charges (\ref{series-entropy}) is shown in  Appendix \ref{app-B}.

Using (\ref{l1l2}) the partition function (\ref{en-chem}) reads as (with $\sigma=9/16$)  
\begin{equation}
\log Z_\gr = \frac{\pi i c}{12 \tau} \lb  1+ \frac{\alpha_2^2}{\tau^2}+\frac{\alpha_3^2}{\tau^4}+ \cdots    \rb
\end{equation}
\def\cft{{\text{CFT}}}
\def\h{{\mathcal{H}}}
\subsection{Partition function from the dual CFT}

We shall now calculate the partition function from the dual CFT. As was noted earlier the bosonic part of  $\mathcal{N}=2$ super-$\cW_3$ algebra  splits as $\cW_+ \oplus \cW_- \oplus u(1) $. The expression for the partition function is then given by
\begin{equation}
Z_\cft = \tr_\h \left[  e^{2\pi i \left(\hat \eta_+ (T^+_0 +\tfrac{1}{2c}J_0^2 - \tfrac{c^+}{24}) +\hat  \eta_- (T^-_0 -\tfrac{2}{c}J_0^2  - \tfrac{c^-}{24}  )     +\hat  \alpha_3 (W_0 - \tfrac{6}{c}(JV)_0) + \hat \alpha_1 J_0  \right)    }				 \right]
\end{equation}
Note that this is the natural CFT partition function one would write down since the 
$\mathcal{N}=2$ super-$\cW_3$ algebra  splits as $\cW_+ \oplus \cW_- \oplus u(1) $.
The partition function simply evaluates the number of states weighted with the charges of 
decoupled bosonic currents. 
The chemical potentials appearing here are those in the CFT and are not the same as those in gravity. 
We would now like to write the CFT partition function in terms of the chemical potentials
appearing the bulk. 
The relation between the two sets of chemical potential is given in   Table 1. 
Using the relations in Table 1  and similar 
definitions as given in (\ref{chem1}) and (\ref{l1l2})  for the chemical 
potentials in the CFT we can obtain the  the following identity
\begin{equation}\label{brel}
\hat \eta_+ (T^+_0 +\tfrac{1}{2c}J_0^2) +\hat  \eta_- (T^-_0 -\tfrac{2}{c}J_0^2) + \hat \alpha_1 J_0 =
  \eta_+ (T^+_0 +\tfrac{1}{2c}J_0^2) +   \eta_- (T^-_0 -\tfrac{2}{c}J_0^2) + \alpha_1 J_0
\end{equation}
This identity allows the rewriting of the CFT partition function in terms of the chemical potentials
of the bulk.  Note that this non-trivial identity was arrived at by using the relations 
given in Table 1 which resulted from the structure of the 
${\cal N}=2$ super ${\cal W}_3$ algebra.  Here again  supersymmetry 
played a role in this crucial simplification.  
The partition function then becomes
\begin{equation}
Z_\cft = q^{-c/24} \tr _\h \left[  e^{2\pi i \left(  \eta_+ (T^+_0 +\tfrac{1}{2c}J_0^2 ) +  \eta_- (T^-_0 -\tfrac{2}{c}J_0^2 ) + \alpha_3 (W_0 - \tfrac{6}{c}(JV)_0) +  \alpha_1 J_0  \right)    }				 \right]
\end{equation}
Since the \(\cW_+, \ \cW_- \) and \(u(1)\) are mutually commuting subalgebras, we have
\begin{equation}
Z_\cft = q^{-c/24} \tr_\h \left[  e^{2\pi i   \eta_+ (T^+_0 +\tfrac{1}{2c}J_0^2 )}    e^{2\pi i \left( \eta_- (T^-_0 -\tfrac{2}{c}J_0^2 ) + \alpha_3  (W_0 - \tfrac{6}{c}(JV)_0)  \right) }   e^{2\pi i  \alpha_1 J_0     }				 \right]
\end{equation}
and also the full Hilbert space of states in the CFT factorizes as
\begin{equation}
\h = \h_+ \oplus \h_- \oplus \h_{u(1)}
\end{equation}
The $u(1)$ part does not contribute to the partition function for our case since we are restricting 
ourselves to the zero charge sector as seen in (\ref{four}). 
The partition function can then be written as contributions from the Virasoro ($\cW_+$) and the $\cW_3$ ($\cW_-$) parts.
\begin{equation}
Z_\cft = q^{-c/24}\  \tr_{\h_+} \left[  e^{2\pi i   \eta_+ (T^+_0 +\tfrac{1}{2c}J_0^2 )} \right] \ \tr_{\h_-} \left[  e^{2\pi i \left( \eta_- (T^-_0 -\tfrac{2}{c}J_0^2 ) + \alpha_3 (W_0 - \tfrac{6}{c}J_0V_0) \right) }  \right] 
\end{equation}
We shall calculate the partition function at high temperature regime $1/ \tau \rightarrow 0$. The leading term for the Virasoro part is 1.  Plugging in the answer calculated for the $\cW_3$ part from \cite{Gaberdiel:2012yb} (with $\lambda=3$, $c\rightarrow c^-$, $\tau\rightarrow\eta_-$ and $\alpha\rightarrow\alpha_3$)\footnote{\cite{Kraus:2011ds, Gaberdiel:2012yb} has $\sigma$ set  to $-1$.}, we get
\begin{equation}\label{zcft}
Z_\cft \simeq  q^{-c/24}  \lb 1+ \frac{2\pi i c^-}{\eta_-} \lb  \frac{\sigma}{18}\frac{\alpha_3^2}{\eta_-^4} - \frac{20\sigma^2}{81}\frac{\alpha_3^4}{\eta_-^8}+\cdots \rb  \rb
\end{equation}
The above answer for the partition function matches precisely with the one calculated from gravity (\ref{Z-bh}). 
It can also be seen easily that (\ref{zcft}) reduces to the Cardy's formula for the BTZ black hole embedded in the gravitational $sl(2)$ when $\alpha_1=0=\alpha_2=\alpha_3$.

\subsection{Generalizations to  higher spin black holes in shs{[$\lambda$]}}

One can generalize the above observations
found for black holes in the $sl(3|2)$ theory to the case of $sl(N|N-1)$ theory 
and the Chern-Simons theory based on the superalgebra shs{[$\lambda$]}. 
The crucial observation which made it possible to obtain the partition function of the 
black hole from the CFT was the existence of charges by which the CFT decoupled into 
three bosonic sub-algebras. The charges of the black hole could  then be mapped to 
charges in each of the sub-algebras. The black hole we studied turned out to have 
higher spin charge in the $sl(3)$ sub-algebra as well as a charge in the $sl(2)$ part. 
Thus appealing to the result of \cite{Gaberdiel:2012yb} and the 
Cardy formula the entropy of the black hole 
was reproduced from the CFT. 

Let us now examine the situation for the Chern-Simons theory based on 
the superalgebra shs{[$\lambda$]}. It is dual to the Kazama-Suzuki model 
based on the coset \cite{Creutzig:2011fe}
\begin{equation}\label{coset}
\frac{SU(N+1)_k \times SO(2N)_1}{SU(N)_{k+1} \times U(1)_{N(N+1)(k+N+1)} }
\end{equation}
The model has a ${\cal W}_{N+1}$ superalgebra whose global part is $sl(N+1|N)$. 
The t' Hooft limit of this coset with 
\begin{equation}
\lambda = \frac{N}{k+N}
\end{equation}
is dual to the shs{[$\lambda$]} Chern-Simons theory. 
As discussed in \cite{Blumenhagen:1994wn, Creutzig:2011fe} the coset admits the following decomposition
\begin{equation} \label{bode}
\frac{SU(N+1)_k \times SO(2N)_1}{SU(N)_{k+1} \times U(1)_{N(N+1)(k+N+1)} } 
\sim \frac{SU(k)_N \times SU(k)_1}{SU(k)_{N+1}} 
\times \frac{SU(N)_k \times SU(N)_1}{SU(N)_{k+1}} \times U(1) 
\end{equation}
These cosets are precisely  of the form
considered by \cite{Gaberdiel:2012uj} in the minimal model/higher spin duality. 
Thus the  bosonic part of the ${\cal W}_{N+1}$ superalgebra decomposes into the algebra
${\cal W}_{\infty} (1-\lambda) \oplus {\cal W}_\infty( \lambda) \oplus u(1) $. It's worthwhile noting that the coset is invariant under the level-rank exchange $N\longleftrightarrow k$ (or $\lambda \longleftrightarrow 1- \lambda$). These theories therefore have a strong-weak self-duality. 
Now any black hole considered in the shs{[$\lambda$]} theory will carry a specific set of 
bosonic charges. The relation (\ref{bode}) implies that, there exists a re-definition
of the charges such that they correspond to charges in the decoupled bosonic 
sub-algebra  of the superalgebra.  Note that the  decomposition in  (\ref{bode}) 
is a property of these ${\cal N}=2$ supersymmetric minimal models, the structure of the super-algebra is important for this decomposition just as we have seen 
in our study of the  ${\cal N}=2$ super ${\cal W}_3$ algebra. 
The partition function of black holes carrying a specific set of higher spin charges constructed by 
\cite{Kraus:2011ds} in the ${\cal W}_\infty(\lambda)$ theory has been reproduced in the CFT 
by the computation done in \cite{Gaberdiel:2012yb}. Therefore we conclude that  the partition function of 
higher spin black holes  constructed by \cite{Kraus:2011ds} embedded in 
${\cal W}_{\infty} (1-\lambda) \oplus {\cal W}_\infty( \lambda) $ algebra will be 
reproduced in the CFT. It will be interesting to find the re-definitions of the charges of the 
black hole so that they can be mapped to charges in the decoupled algebras for the 
shs{[$\lambda$]} just as we have done for the $sl(3|2)$ theory. 
To find these redefinitions one will have to investigate the complete OPE's of the theory 
which is constrained by supersymmetry.

\section{Conclusions}
In this paper we have obtained the relations between the 
definitions of the charges and the chemical potentials between the boundary CFT and 
the bulk $sl(3|2)$ Chern-Simons theory. These relations are summarized in Table 1. 
From these relations we observed that a natural linear combinations of the bulk charges 
can be identified with the charges of the decoupled bosonic sub-algebras of the 
super-${\cal W}_3$ CFT. 
We then constructed a higher spin black hole in this theory and evaluated both its  entropy
and partition function. The entropy was shown to be a sum of contributions from the decoupled 
sub-algebras. The decomposition of the CFT into decoupled sectors also enabled the evaluation of the 
partition function which was shown to be in precise agreement with that evaluated in the bulk. 

Let us now summarize where supersymmetry plays a role in our analysis. 
Firstly the question of studying the effect of the $U(1)$ charge in Vasiliev 
theories is naturally realizable in theories based on supergroup $sl(N|N-1)$ which 
contains a $U(1)$ as part of the R-symmetry. 
Secondly the shifts in the definition of  the charges and the chemical potentials given in 
Table 1 was  obtained from the detailed analysis the ${\cal N}=2$ super
${\cal W}_3$-algebra. Finally  the fact that this superalgebra decouples as 
in (\ref{decoup1})  and 
the redefinitions of Table 1  were crucial in the simplification
of the CFT partition function which is  summarized in the identity (\ref{brel}). 
This identity related the CFT chemical potentials to that of the bulk which 
enabled the evaluation and the comparison of the partition functions evaluated in the 
boundary CFT and that from the bulk.

As we have discussed these observations can be generalized to the case of Chern-Simons 
theory based on the supergroup shs{[$\lambda$]}. It will be interesting to find the analog of 
the relations given in Table 1 for this case. These relations will facilitate the evaluation
the entropy of the higher spin black hole in these theories. 
Another direction to explore is the implication of the relations in Table 1 and its counterparts for smooth  conical defects in these theories. 
It is possible that these classical solutions can be related to the vacuum.



\section*{Acknowledgements}
We would like to thank Michael Ferlaino, Matthias Gaberdiel,  Rajesh Gopakumar, S.~Prem Kumar and Amitabh Virmani  for useful discussions. S.D.~thanks ETH, Zurich and ICTP, Trieste for hospitality during which a part of this work was completed. 
The work J.R.D.~is partially supported by  the Ramanujan fellowship DST-SR/S2/RJN-59/2009. 
\def\zw{(z-w)}

\appendix

\section{OPEs of $\mathcal{N}=2$ super-$\cW_3$}
Following \cite{Romans:1991wi}, we shall list the operator product expansions which were used to calculate the Ward identities in Section 2. 

\begin{align}
J(z)J(w) &\sim \frac{c/3}{(z-y)^2} \\
J(z)W(w) &\sim \frac{2}{(z-w)^2} V(w)
\end{align}

\begin{align}
T(z)J(w) &\sim \frac{J(w)}{(z-w)^2} + \frac{\pd_w J(w)}{(z-y)} \\
T(z)V(w) &\sim \frac{2V(w)}{(z-w)^2} + \frac{\pd_w V(w)}{(z-y)} \\
T(z)W(w) &\sim \frac{3W(w)}{(z-w)^2} + \frac{\pd_w W(w)}{(z-y)} 
\end{align}

\begin{align}
V(z)V(w) &\sim \frac{2 A^{[2]}}{(z-w)^2} + \frac{\pd_w A^{[2]}}{(z-w)} + \frac{c/2}{(z-w)^4} \\
V(z)W(w) &\sim \frac{1}{(z-w)}C^{[4]} + \lb \frac{3}{(z-w)^2} + \frac{1}{(z-w)} \pd_w \rb C^{[3]} \nn \\
&\qquad + \lb \frac{20}{\zw^3} + \frac{5}{\zw^2}\pd_w + \frac{1}{\zw}\pd_w^2  \rb C^{[2]} + \frac{6}{\zw^4}C^{[1]} \tag*{(\theequation)\footnotemark}
\end{align}  \footnotetext{The last term here had a 36 instead of 6 in \cite{Romans:1991wi} as a possible typographical error. The OPE we have written here gives the correct commutation relation.}

\begin{align}
W(z)J(w) &\sim \frac{2}{\zw^2} V(w) + \frac{2}{\zw}\pd_w V(w) \\
W(z)V(w) &\sim \frac{1}{\zw^3}(-20C^{[2]}+6\pd_w C^{[1]}) + \frac{1}{\zw^2}(3C^{[3]}  - 15\pd_w C^{[2]} +3 \pd_w^2 C^{[1]}) \nn \\
& \quad +\frac{1}{\zw}(-C^{[4]} + 2 \pd_w C^{[3]} - 6\pd_w^2 C^{[2]} + \pd_w^3 C^{[1]} ) + \frac{6}{\zw^4}C^{[1]} \\
W(z)W(w) &\sim \frac{c_W}{3\zw^6} + \lb \frac{2}{\zw^2} + \frac{1}{\zw} \pd_w \rb B^{[4]} \nn \\
& \quad+ \lb    \frac{60}{\zw^4}   + \frac{30}{\zw^3} \pd_w + \frac{9}{\zw^2}\pd_w^2 + \frac{2}{\zw}\pd^2_w   \rb B^{[2]}
\end{align}
The  exact expressions  of $A^{[\; ]},\ B^{[\; ]}$ and $C^{[\;]}$ appearing in the OPEs above  are given in \cite{Romans:1991wi}. However, we are interested in the semiclassical limit of $c\rightarrow\infty$. It can seen from the equations of motion that the operators $\cL,\ \cV$ and $\cW$ scale as $O(c)$, while all the chemical potentials scale as $O(1)$. On retaining just the $O(c)$ terms in the OPEs $A^{[\; ]},\ B^{[\; ]}$ and $C^{[\;]}$ are expressed as follows
\begin{align} \label{abc}
A^{[2]}&=T-\frac{3}{2c}J^2 + \kappa V \nn \\
B^{[2]}&=\frac{1}{4}\lb T+\frac{\kappa}{5}V \rb - \frac{3}{40c} J^2 \nn \\
B^{[4]}&=\frac{24}{c}T^2 - \frac{36}{c^2}J^2 T +\frac{9}{2c}J \pd^2J - \frac{27}{20c}\pd^2(J^2)+\frac{24i}{c}TV+\frac{9i}{c}JW \nn  \\
C^{[1]}&=\frac{1}{2}J \nn \\
C^{[{2}]}&=0 \nn \\
C^{[3]}&=\frac{4}{c}JT - \frac{6}{c^2} J^3 +\frac{i}{2} W + \frac{7i}{c}JV  \\
C^{[4]}&= \frac{2}{c} (J\pd(T+\kappa V + \frac{3}{2c}J^2 ) - 2\pd J (T+\kappa V + \frac{3}{2c}J^2 ) ) \nn
\end{align}
While obtaining the above the fermionic operators were set to zero.

\section{Equivalence of the black hole entropy formulae}
\label{app-B}
We shall prove the equivalence of the formula for the entropy of the higher spin black hole written in terms of charges (\ref{en-gravity}) or (\ref{series-entropy}) with that of the formula in terms of the chemical potentials (\ref{en-chem}). 

The formula for the entropy in terms of the chemical potentials reads as
\begin{align} \label{en-chem2}
S (\eta_+,\eta_-,\alpha_3) =\frac{i \pi  c^+}{6 \eta_+ } + \frac{i \pi  c^-}{6 \eta_- } \lb 1+ \frac{8\sigma}{3} \frac{\alpha_3^2}{\eta_-^4}+ \frac{160\sigma^2}{9}\frac{\alpha_3^4}{\eta_-^4} + \cdots \rb 
\end{align}
We assume a series expansion for $S(\cT_+,\cT_-,\cW_-)$ of the form
\begin{equation} \label{en-charge2}
S(\cT_+,\cT_-,\cW_-)=2 \pi  \sqrt{2 \pi  k_+\cT_+} +2 \pi  \sqrt{2 \pi  k_-\cT_-}     \sum _{j=0}^\infty \chi _j \left(   \frac{\cW_-^2}{\cT_-^3}\right)^j
\end{equation}
One can substitute (\ref{en-charge2}) on the LHS of (\ref{en-chem2}) and find $\alpha_3 = \frac{i}{4\pi^2}\frac{\pd S}{\pd \cW}$ and  $\eta_i = \frac{i}{4\pi^2}\frac{\pd S}{\pd \cL_i}$ and substitute them on the RHS of (\ref{en-chem2}), to get the following equations
\begin{align}
\chi_0 &= 1 \nn \\
256 \pi  \sigma  \chi _1^2 &= 3 k \chi _1   \\
5 \left(45 k^2 \chi _1^2+18 k^2 \chi _2+524288 \pi ^2 \sigma ^2 \chi _1^4\right)&=3072 \pi  k \sigma  \chi _1 \left(25 \chi _1^2+4 \chi _2\right) \nn    \\
&\ \,  \vdots \nn 
\end{align}
which have the solutions 
\begin{equation}
\chi_0 =1\ , \quad \chi_1 =\frac{3 k_-}{256 \pi  \sigma }  \ , \quad  \chi_2 = -\frac{105 k_-^2}{131072 \pi ^2 \sigma ^2} \ , \ \cdots
\end{equation}
This is in precise agreement with the expansion in terms of the charges  given in (\ref{series-entropy}).

\providecommand{\href}[2]{#2}\begingroup\raggedright\endgroup

\end{document}